\documentclass[a4paper,fleqn,usenatbib]{mnras}
\usepackage{txfonts}
\usepackage{graphicx}
\usepackage{upgreek}

\title[Astrometry of the galaxy associated with
  FRB\,150418]{Optical and radio astrometry of the galaxy associated with
  FRB\,150418}

\author[Bassa et al.]  {C.\,G.\,Bassa$^1$\thanks{email: bassa@astron.nl},
  R.\,Beswick$^{2}$,
  S.\,J.\,Tingay$^{3,4}$,
  E.\,F.\,Keane$^{5,6,7}$,
  S.\,Bhandari$^{6,7}$,
  S.\,Johnston$^{8}$,\newauthor
  T.\,Totani$^{9}$,
  N.\,Tominaga$^{10,11}$,
  N.\,Yasuda$^{11}$,
  B.\,W.\,Stappers$^{2}$,
  E.\,D.\,Barr$^{6}$,
  M.\,Kramer$^{12,2}$,\newauthor
  A.\,Possenti$^{13}$\newauthor\\
  $^{1}$ASTRON, the Netherlands Institute for Radio Astronomy, Postbus 2, NL-7990 AA Dwingeloo, The Netherlands\\
  $^{2}$Jodrell Bank Centre for Astrophysics, School of Physics and Astronomy, University of Manchester, Manchester M13 9PL, UK\\
  $^{3}$International Centre for Radio Astronomy Research (ICRAR), Curtin University, Bentley, WA 6102, Australia\\
  $^{4}$Istituto Nazionale di Astrofisica (INAF) -- Istituto di Radio Astronomia, Via Piero Gobetti, Bologna, 40129, Italy\\
  $^{5}$Square Kilometre Array Organisation, Jodrell Bank Observatory, SK11 9DL, UK\\
  $^{6}$Centre for Astrophysics and Supercomputing, Swinburne University of Technology, Mail H29, PO Box 218, Victoria 3122, Australia\\
  $^{7}$Australian Research Council Centre of Excellence for All-sky Astrophysics (CAASTRO), Australia\\
  $^{8}$CSIRO Astronomy and Space Science, Australia Telescope National Facility, PO Box 76 Epping, NSW, 1710, Australia\\
  $^{9}$Department of Astronomy, the University of Tokyo, Hongo, Tokyo 113-0033, Japan\\
  $^{10}$Department of Physics, Faculty of Science and Engineering, Konan University, 8-9-1 Okamoto, Kobe, Hyogo 658-8501, Japan\\
  $^{11}$Kavli Institute for the Physics and Mathematics of the Universe (WPI), Institutes for Advanced Study, University of Tokyo, Kashiwa, Chiba 277-8583, Japan\\
  $^{12}$Max-Planck-Institut f\"ur Radioastronomie (MPIfR), Auf dem H\"ugel 69, D-53121 Bonn, Germany\\
  $^{13}$Istituto Nazionale di Astrofisica (INAF)-Osservatorio Astronomico di Cagliari, Via della Scienza 5, I-09047 Selargius (CA), Italy\\
}

\date{Accepted 27 July 2016. Received 2016 July 5; in original form 2016 May 12}

\pubyear{2016}

\begin{document}
\label{firstpage}
\pagerange{\pageref{firstpage}--\pageref{lastpage}} 
\maketitle

\begin{abstract}
  A fading radio source, coincident in time and position with the fast
  radio burst FRB\,150418, has been associated with the galaxy
  WISE\,J071634.59$-$190039.2. Subsequent observations of this galaxy
  have revealed that it contains a persistent, but variable, radio
  source. We present e-MERLIN, VLBA, and ATCA radio observations and
  Subaru optical observations of WISE\,J071634.59$-$190039.2 and find
  that the persistent radio source is unresolved and must be compact
  ($<0.01$\,kpc), and that its location is consistent with the optical
  centre of the galaxy. We conclude that it is likely that
  WISE\,J071634.59$-$190039.2 contains a weak radio AGN.
\end{abstract}


\section{Introduction}
Fast radio bursts (FRBs, see e.g.~\citealt{pbj+16} and references
therein) are millisecond-duration bursts of radio emission that have
been observed at the Parkes, Arecibo, and Green Bank radio
telescopes~(\citealt{lbm+07,sch+14,mls+15}). FRBs have dispersion
measures (DMs), a measure of the electron column density, that range
from $1.4$ to $33$ times the maximum Galactic
contribution~\citep{cl02}, thought to be attributable to free
electrons in the intergalactic medium. With this interpretation the
distances to FRBs are cosmological~\citep{lbm+07,tsb+13}, and the
corresponding luminosities of the FRB signals are thus many orders of
magnitude higher than typical pulsar luminosities.

Non-cosmological explanations have been put forward
(e.g.\,\citealt{bbe+11,lsm14,kon+14}) but a cosmological
interpretation remains favored, based on current observational
evidence. While the extragalactic interpretation of FRBs currently
prevails, their progenitor(s) are as yet unknown. In an effort to
determine the nature of FRBs, the SUrvey for Pulsars and Extragalactic
Transients (SUPERB) performs real time FRB searches at the Parkes
telescope, and employs an array of multi-wavelength telescopes to
follow up FRB discoveries. Multi-wavelength follow-up of FRB\,150418
led, for the first time, to the detection of a fading radio source
that was associated with a galaxy at $z=0.49$~\citep{kjb+16}. This
galaxy is also detected in the mid-infrared by WISE \citep{wem+10} and
cataloged as WISE\,J071634.59$-$190039.2.

Radio imaging observations of WISE\,J071634.59$-$190039.2 with the
Australia Telescope Compact Array (ATCA) showed a source declining by
a factor $\sim3$ in brightness at 5.5\,GHz during the first 6\,d after
the FRB. This source subsequently settled at a persistent brightness
of approximately 100\,$\upmu$Jy\,beam$^{-1}$. Comparing this behaviour
to the results of transient surveys \citep{bhh+15,mhb+16} led
\citet{kjb+16} to argue in favour of the association between
FRB\,150418 and the fading radio source, and hence with the galaxy.

The association of FRB\,150418 with the fading radio source and hence
with WISE\,J071634.59$-$190039.2 met with criticism. JVLA observations
of the persistent radio source in WISE\,J071634.59$-$190039.2 showed
rapid variability, which led \citet{wb16} to argue that the
variability of the fading radio source is consistent with the
intrinsic or scintillating behaviour of a compact, weak active
galactic nucleus (AGN). Similarly, \citet{aj16} show that the
variability of the persistent source may be extrinsic and attributable
to refractive scintillation in the Milky Way, assuming a compact radio
source is present in the galaxy. \citet{vrm+16} find a flat radio
spectrum for the persistent source and suggest it is consistent with
the properties of an AGN.

Since FRB\,150418 is the first FRB for which a radio counterpart and
host galaxy have been suggested, the host galaxy warrants closer
study. In this Letter we report on an astrometric radio and optical
analysis that establishes that WISE\,J071634.59$-$190039.2 currently
contains a single weak, compact radio source consistent with an AGN
located at the centre of the galaxy.

\section{Observations and Analysis}\label{sec:observations}

\subsection{ATCA}
We observed WISE\,J071634.59$-$190039.2 with the ATCA on 2016 March 1
at 05:30 UTC for a duration of 11\,hours. Observations were made in 6B
configuration, in two frequency bands each with bandwidth 2\,GHz
centered at 5.5 and 7.5\,GHz respectively. Single pointing mode was
used which yielded an image rms of 6\,$\upmu$Jy\,beam$^{-1}$ with an
angular resolution of $1\farcs9\times10\farcs4$. Bandpass and flux
density calibration were carried out using the standard ATCA
calibrators B1934$-$638 and B0823$-$500 and phase calibration with
B0733$-$174. Data reduction was performed in MIRIAD \citep{stw95}
using standard techniques and the position of sources in the field
were measured using the task IMFIT. The position is listed in
Table\,\ref{tab:sources}. In addition, we used the mosaic observations
made with the ATCA on 2015 October 27 and described in \citet{kjb+16}
to measure the positions of four other sources in the field
(Table\,\ref{tab:othersources}).

\subsection{e-MERLIN}
WISE\,J071634.59$-$190039.2 was observed with the e-MERLIN array on
2016 March 18, 21 and 22 between 17:00 and 21:50 UTC on each day at
C-band (4.816--5.328\,GHz). The e-MERLIN array comprised of six
telescopes for each of these observations. The 76-m Lovell telescope
was not included. Data were correlated in full Stokes mode in a
standard C-band configuration with $4\times128$\,MHz bands. The target
and phase calibrator (J0718$-$1813) were observed with a 7:3 minute
phase referencing cycle. In addition, every hour, one target scan was
re-deployed to observe a faint nearby NVSS radio source (J0716$-$1908)
in order to verify calibration. Observations of 3C286 and OQ208
(30\,mins each) were made for flux density and bandpass calibration.

Each observing run was reduced independently and in an identical
manner. Data were imported into AIPS \citep{gre03} using the e-MERLIN
pipeline \citep{argo15}. Following editing, these data were
fringe-fitted for delay only and flux density calibrated using 3C286
relative to the \citet{pb13} flux density scale. Bandpass solutions
were derived using the point source calibrator OQ208. Standard phase
referencing calibration was applied using the nearby phase calibrator
source J0718$-$1813.  These phase and amplitude calibration solutions
were then applied to both the target and check calibration source
(J0716$-$1908) which were subsequently imaged using standard
techniques.

Following initial imaging of individual epochs, data from all three
observing runs were concatenated together to increase
sensitivity. These data were then re-weighted in a time and frequency
dependent manner to maximize the sensitivity. An unresolved radio
source associated with WISE\,J071634.59$-$190039.2 was detected with a
measured peak brightness of 151 $\upmu$Jy\,beam$^{-1}$ and an image
rms of 21 $\upmu$Jy\,beam$^{-1}$. The angular resolution is
$0\farcs251\times0\farcs030$ at a position angle of $11\fdg8$. In
addition to the target source, two in-beam radio sources were
detected. The coordinates of these sources and
WISE\,J071634.59$-$190039.2 are listed in Tables\,\ref{tab:sources}
and \ref{tab:othersources}.

In addition to these C-band observations, e-MERLIN also observed at
L-band (1.25--1.75\,GHz) on 2016 Apr 2 between 16:15 and
21:00\,UTC. These observations utilized all e-MERLIN telescopes
including the Lovell telescope. Data were correlated across 512\,MHz
of bandwidth and divided in to 8 bands. The same phase calibration
source (J0718$-$1813) as employed at C-band was used, and data were
reduced in an identical manner.

Following wide field imaging of the target field at L-band, e-MERLIN
detected 7 radio sources within the wider field of view accessible to
the array. The positions of these sources are listed in
Table\,\ref{tab:othersources}. Radio emission at the location of the
target source, WISE\,J071634.59$-$190039.2, was detected at a peak
flux density of $92\pm19$\,$\upmu$Jy\,beam$^{-1}$. The angular
resolution is $0\farcs46\times0\farcs32$ at a position angle of
$-6\fdg6$.

\subsection{VLBA}
The Very Long Baseline Array (VLBA; \citealt{nbc+94}) was used to
observe WISE\,J071634.59$-$190039.2 on 2016 March 8, 00:30--06:30\,UTC
at C-band (4.852--5.108\,GHz). Data were recorded in dual circular
polarization mode for 8$\times$32\,MHz bands with 2-bit Nyquist
sampling, corresponding to an aggregate recorded data rate of
2048\,Mbps. All ten antennas of the VLBA were used. As well as the
target galaxy, two phase reference sources were observed, J0719$-$1955
and J0718$-$1813 (same phase calibrator as used for
e-MERLIN). Observations of each individual phase reference source were
made once per 10 minutes, but interleaved such that a calibrator was
observed every 5 minutes.

The data were correlated using the DiFX software correlator
\citep{dtbw07,dbp+11} using an integration time of 1.024\,seconds and
with 64 frequency channels per 32\,MHz of bandwidth.  As well as
correlation phase centers set on the target galaxy and phase reference
positions, additional phase centers (J2000) were correlated, for
sources known from our ATCA imaging within the VLBA primary beam while
pointed at the target galaxy.  Four additional phase centers were
correlated ($\alpha_\mathrm{J2000}$, $\delta_\mathrm{J2000}$):
($07^\mathrm{h}16^\mathrm{m}39\fs41$, $-18\degr56\arcmin29\farcs2$);
($07^\mathrm{h}16^\mathrm{m}04\fs04$, $-19\degr00\arcmin14\farcs7$);
($07^\mathrm{h}16^\mathrm{m}05\fs47$, $-19\degr00\arcmin15\farcs9$);
and ($07^\mathrm{h}16^\mathrm{m}14\fs41$,
$-19\degr06\arcmin49\farcs9$).

The correlated data were processed in AIPS \citep{gre03} using
standard phase reference techniques. System temperature and gain
information for each antenna was used to calibrate the visibility
amplitudes, in addition to corrections for sampler
thresholds. Corrections for instrumental effects, Earth orientation
parameters, and ionosphere were applied to the visibility phases. The
data for the two phase reference calibrators were then fringe-fitted,
exported from AIPS, and imaged in DIFMAP \citep{spt95}. The images
were imported to AIPS and used as source brightness models in a second
round of fringe-fitting, to account for their structure contributions
to the visibility phases. The resultant phase corrections were
smoothed over a 0.4-hour period using boxcar averaging. The final
aggregate amplitude and phase corrections were transferred to the
correlated data for each of the target and in-beam radio sources and
applied to the visibilities before writing the data to disk as FITS
files, retaining full frequency channelization.

The FITS files were read in to MIRIAD and naturally weighted Stokes I
images ($\pm1\farcs5$ around the phase centre position) were produced,
using multi-frequency synthesis to avoid bandwidth smearing, with
0.3\,mas\,pix$^{-1}$ sizes (10,000 pixel images). The target radio
source associated with WISE\,J071634.59$-$190039.2 was detected with a
peak brightness of 130\,$\upmu$Jy\,beam$^{-1}$ and an image rms of
14\,$\upmu$Jy\,beam$^{-1}$. The position, obtained using task MAXFIT
and checked with task IMFIT, is listed in
Table\,\ref{tab:sources}. The uncertainty on the position includes the
uncertainty on the position of the phase reference calibrators (< 1
mas in both coordinates) as given by \citet{pkfg06}. No evidence of
resolved emission is seen and the object appears point-like
($<1.5$\,mas in extent). The angular resolution (full width at half
maximum of the point spread function) of the VLBA images is
$0.9\,\mathrm{mas}\times2.6\,\mathrm{mas}$ at a position angle of
$-3\fdg4$.  Two of the in-beam sources were also detected, and
coincide with e-MERLIN and ATCA sources 1 and 2 (see
Table\,\ref{tab:othersources}).

\begin{table}
  \caption{Radio and optical positions for
    WISE\,J071634.59$-$190039.2. The uncertainties of the Subaru
    positions have the uncertainty in the astrometric calibration
    added in quadrature. }
  \label{tab:sources}
  \begin{tabular}{llll}
    \hline\hline
    Telescope & Band & $\alpha_\mathrm{J2000}$ & $\delta_\mathrm{J2000}$ \\
    \hline
    e-MERLIN &   L & $07^\mathrm{h}16^\mathrm{m}34\fs554(2)$  & $-19\degr00\arcmin39\farcs42(5)$ \\
    e-MERLIN &   C & $07^\mathrm{h}16^\mathrm{m}34\fs5550(4)$ & $-19\degr00\arcmin39\farcs466(18)$ \\
    VLBA     &   C & $07^\mathrm{h}16^\mathrm{m}34\fs5550(1)$ & $-19\degr00\arcmin39\farcs476(3)$ \\
    ATCA &       C & $07^\mathrm{h}16^\mathrm{m}34\fs573(10)$ & $-19\degr00\arcmin38\farcs6(7)$ \\
    Subaru   & $i^\prime$ & $07^\mathrm{h}16^\mathrm{m}34\fs556(7)$    & $-19\degr00\arcmin39\farcs53(8)$ \\
    \hline
  \end{tabular}
\end{table}

\begin{figure*}
  \includegraphics[width=\columnwidth]{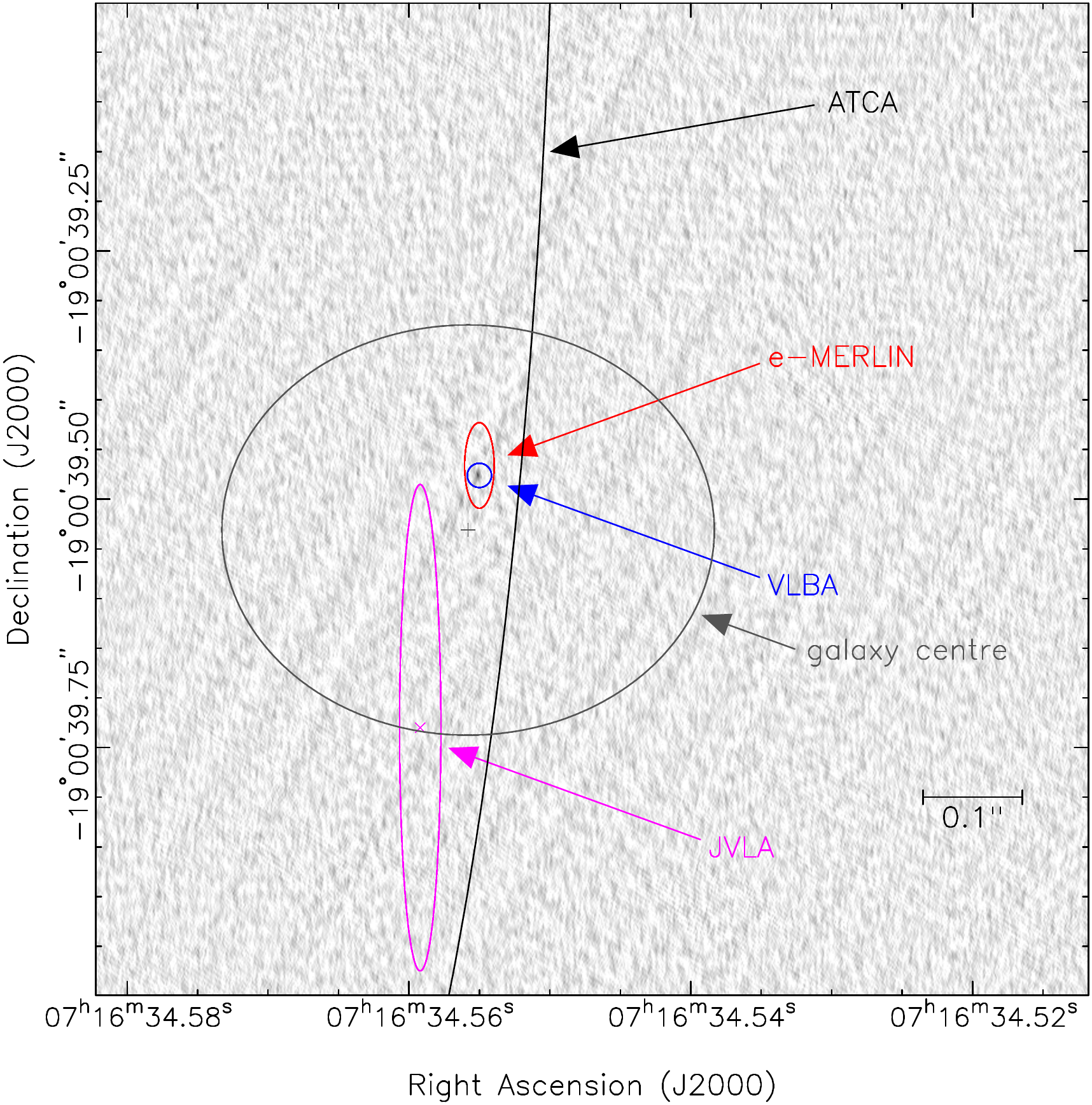}
  \includegraphics[width=\columnwidth]{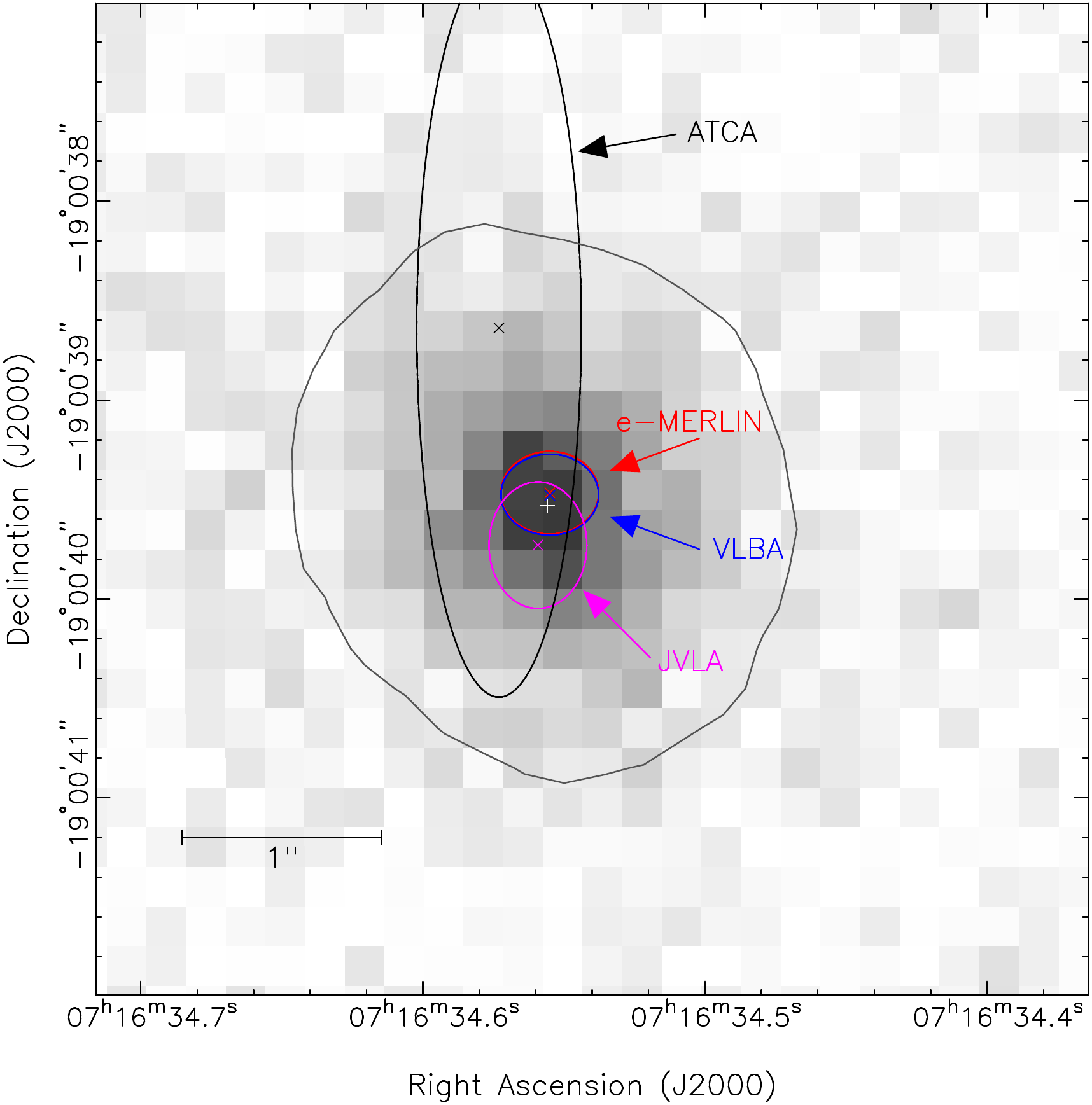}
  \caption{The right panel shows a $1\arcsec\times1\arcsec$ subsection
    of the VLBA radio image of the galaxy centre. Here, the position
    and uncertainty of the optical centre of the galaxy are shown with
    the plus sign and the ellipse (95\% confidence). The left panel
    shows a $5\arcsec\times5\arcsec$ subsection of the 600\,s Subaru
    Supreme-Cam $i^\prime$-band image of the
    WISE\,J071634.59$-$190039.2 galaxy. The centre of light of the
    galaxy on the image is marked by the white plus sign. The dark
    grey contour traces the half-light radius as defined by
    \citet{kjb+16}. In both panels, the 95\% confidence uncertainties
    on the positions of the radio source seen by e-MERLIN, VLBA, ATCA
    and JVLA \citep{vrm+16} are shown with the ellipses. Note that in
    the right panel, these include the uncertainty in the astrometric
    calibration of the Subaru image, hence the differences in error
    region sizes between the two panels.}
  \label{fig:chart}
\end{figure*}

\begin{table}
  \caption{Radio and optical positions 7 sources in the
    field-of-view. The uncertainties of the Subaru positions have the
    uncertainty in the astrometric calibration added in quadrature. }
  \label{tab:othersources}
  \begin{tabular}{llll}
    \hline\hline
    Telescope & Band & $\alpha_\mathrm{J2000}$ & $\delta_\mathrm{J2000}$ \\
    \hline
    \multicolumn{4}{l}{Source 1 (NVSS\,J071639$-$185620)}\\
    \hline
    e-MERLIN &   L & $07^\mathrm{h}16^\mathrm{m}39\fs408(2)$ & $-18\degr56\arcmin29\farcs93(3)$ \\
    e-MERLIN &   C & $07^\mathrm{h}16^\mathrm{m}39\fs4080(5)$ & $-18\degr56\arcmin29\farcs89(2)$ \\
    VLBA     &   C & $07^\mathrm{h}16^\mathrm{m}39\fs4082(1)$    & $-18\degr56\arcmin29\farcs890(3)$ \\
    ATCA &       C & $07^\mathrm{h}16^\mathrm{m}39\fs410(2)$ & $-18\degr56\arcmin29\farcs30(19)$ \\
    Subaru   & $i^\prime$ & $07^\mathrm{h}16^\mathrm{m}39\fs408(7)$ & $-18\degr56\arcmin30\farcs06(9)$ \\
    \hline
    \multicolumn{4}{l}{Source 2 (NVSS\,J071604$-$190015)}\\
    \hline
    e-MERLIN &   L & $07^\mathrm{h}16^\mathrm{m}04\fs037(2)$ & $-19\degr00\arcmin15\farcs78(3)$ \\
    e-MERLIN &   C & $07^\mathrm{h}16^\mathrm{m}04\fs0370(6)$  & $-19\degr00\arcmin15\farcs86(3)$ \\
    VLBA     &   C & $07^\mathrm{h}16^\mathrm{m}04\fs0355(1)$   & $-19\degr00\arcmin15\farcs775(3)$ \\
    ATCA &       C & $07^\mathrm{h}16^\mathrm{m}04\fs034(1)$ & $-19\degr00\arcmin14\farcs79(15)$ \\
    Subaru   & $i^\prime$ & $07^\mathrm{h}16^\mathrm{m}04\fs030(7)$  & $-19\degr00\arcmin15\farcs83(8)$ \\
    \hline
    \multicolumn{4}{l}{Source 3}\\
    \hline
    e-MERLIN &   L & $07^\mathrm{h}16^\mathrm{m}05\fs475(2)$ & $-19\degr00\arcmin16\farcs88(3)$ \\
    ATCA &       C & $07^\mathrm{h}16^\mathrm{m}05\fs470(4)$ & $-19\degr00\arcmin15\farcs9(4)$ \\
    Subaru   & $i^\prime$ & $07^\mathrm{h}16^\mathrm{m}05\fs467(8)$ &  $-19\degr00\arcmin16\farcs78(10)$ \\
    \hline
    \multicolumn{4}{l}{Source 4}\\
    \hline
    e-MERLIN & L & $07^\mathrm{h}16^\mathrm{m}14\fs405(2)$ & $-19\degr06\arcmin50\farcs47(4)$ \\
    ATCA &       C & $07^\mathrm{h}16^\mathrm{m}14\fs403(7)$ & $-19\degr06\arcmin49\farcs8(7)$ \\
    \hline
    \multicolumn{4}{l}{Source 5}\\
    \hline
    e-MERLIN & L & $07^\mathrm{h}16^\mathrm{m}02\fs558(3)$ & $-19\degr08\arcmin19\farcs89(4)$ \\
    \hline
    \multicolumn{4}{l}{Source 6}\\
    \hline
    e-MERLIN & L & $07^\mathrm{h}16^\mathrm{m}19\fs356(3)$ & $-19\degr13\arcmin56\farcs94(4)$ \\
    \hline
    \multicolumn{4}{l}{Source 7}\\
    \hline
    e-MERLIN & L & $07^\mathrm{h}16^\mathrm{m}47\fs200(3)$ & $-18\degr45\arcmin25\farcs26(5)$ \\
    \hline
  \end{tabular}
\end{table}

\subsection{Subaru}
Optical observations of the field containing FRB\,150418 were obtained
with Suprime-Cam on the 8.2-m Subaru telescope. Here, we use the
$i^\prime$-band observation obtained on 2015, April 19 with a dithered
set of $10\times60$\,s exposures. Suprime-Cam is a mosaic of ten
4k$\times$2k detectors, covering a field-of-view of $34\arcmin \times
27\arcmin$ sampled at $0\farcs2$\,pix$^{-1}$. The individual images
were reduced using the Hyper-Suprime-Cam pipeline version 3.8.5, which
is developed based on the LSST pipeline \citep{ita+08,aklp10}.
Following the bias subtraction and flatfielding, the astrometry of all
images is simultaneously solved with a 5th order polynomial to
represent the optical distortion. Here we used 4606 stars for the
fitting to the external catalog and 3826 stars for the internal
fitting. The weights of images for co-adding them are also derived
according to their signal-to-noise ratios. Thereafter, the individual
images were mapped to pixels on a world coordinate system and a
weighted mean of each pixel value was computed with clipping of
$3\sigma$ outliers.

To calibrate the astrometry of this co-added image, we compared the
centroids of stars on a $14\arcmin \times 14\arcmin$ subsection with
several astrometric catalogs. We selected only bright ($K<14$) stars
from the 2MASS catalog \citep{csd+03,scs+06} that matched objects on
the Suprime-Cam image that were not saturated and appeared stellar and
unblended. Iteratively rejecting outliers with residuals in excess of
$0\farcs25$, the final astrometric calibration, fitting for zero-point
position, and a four parameter transformation matrix, uses 88 stars
and yields rms residuals of $0\farcs058$ in right ascension and
$0\farcs060$ in declination.

Since the 2MASS catalog does not provide proper motions, the
calibration might have a systematic positional offset due to
non-random proper motions over the $\sim15$\,yr time baseline between
the 2MASS observations and the Subaru observation. To investigate the
effect of proper motion we also calibrated the same subsection of the
Suprime-Cam image against the 4th version of the USNO CCD Astrograph
Catalog (UCAC4; \citealt{zfg+13}). This catalog provides proper
motions by combining positional CCD measurements obtained between 1998
and 2004 with positions from historic astrograph plates. Propagating
the UCAC4 positions to the epoch of the Subaru observations, and
applying the same procedure as with 2MASS, we obtain an astrometric
calibration using 78 stars and rms residuals of $0\farcs101$ in right
ascension and $0\farcs083$ in declination. 

Based on the UCAC4 calibration, the centre of light of
WISE\,J071634.59$-$190039.2 is offset from the position based on the
2MASS calibration by $0\farcs047$ in right ascension and $0\farcs025$
in declination. Since the UCAC4 calibration corrects for proper motion
and encloses the positional uncertainty of WISE\,J071634.59$-$190039.2
based on the 2MASS calibration, we use the UCAC4 calibration for the
remainder of the paper. Of the 7 additional radio sources detected by
e-MERLIN at L-band, 6 overlap with the full Subaru image, and 3 have
optical counterparts above the $5\sigma$ limit of $i^\prime=24.7$. The
positions of these counterparts, as well as that of
WISE\,J071634.59$-$190039.2 are given in Table\,\ref{tab:sources} and
\ref{tab:othersources}. The optical positions of these counterparts
are consistent with the radio positions, providing independent
confirmation that the astrometric calibration is correct.

\section{Discussion}\label{sec:discussion}
A single unresolved radio source is detected in the ATCA, e-MERLIN and
VLBA observations of WISE\,J071634.59$-$190039.2. The source has a
brightness of 130 to 151\,$\upmu$Jy\,beam$^{-1}$ at frequencies
between approximately 4.8 and 5.3\,GHz, consistent within the
uncertainties. The coordinates derived from the ATCA, e-MERLIN and
VLBA observations are in agreement and are consistent with previously
reported ATCA \citep{kjb+16} coordinates. These detections are
consistent with the low significance detection at the same position
with the EVN, reported by \citet{mgg+16a,mgg+16b}, after our high
resolution astrometry was first reported as an ATel
\citep{bbt+16a}. We note that the JVLA position by \citet{vrm+16} is
inconsistent in right ascension with our e-MERLIN and VLBA positions.

The coordinates of the compact radio source are plotted in
Figure\,\ref{fig:chart} and are overlaid on the Subaru Suprime-Cam
$i^\prime$-band image of WISE\,J071634.59$-$190039.2 that was obtained
on 2015, April 19, as well as the VLBA radio image. We conservatively
plot 95\% confidence uncertainty regions of the radio source
positions. The optical centre of light of WISE\,J071634.59$-$190039.2
is offset from the C-band VLBA and e-MERLIN positions by $\Delta
\alpha=0\farcs01(10)$ and $\Delta \delta=-0\farcs05(8)$. On this
basis, our astrometry shows that, within the uncertainties quoted
above, the location of the compact radio source is consistent with the
optical centre of light of WISE\,J071634.59$-$190039.2 in the Subaru
$i^\prime$-band image.

The compact radio source detected recovers a very high percentage of
the persistent flux density reported by \citet{kjb+16} and
\citet{vrm+16}, indicating that little, if any, extended radio
emission exists.  The upper limit to the size of the radio source of
$<1.5$\,mas (implying a brightness temperature in excess of
$5\times10^6$\,K), corresponds to a physical size of less than
0.01\,kpc at a redshift of $z=0.49$.  An interpretation of the radio
emission as due to a star formation region therefore appears highly
unlikely as circumnuclear star formation regions (e.g. NGC\,253;
\citealt{lt06}), those associated with merger activity \citep{edg+11},
or jet induced star formation regions \citep{ssc14}, are generally two
orders of magnitude larger than our upper limit. Furthermore, the
observed brightness temperature exceeds that expected from thermal
radio emission processes associated with star formation regions.  The
location of the compact radio source is consistent with our best
estimate of the centre of its host galaxy.  Thus, our data are
consistent with the existence of a weak radio AGN within the galaxy
(see \citealt{gbp+13} for a discussion of the existence of AGN in
`radio quiet' galaxies).

While the emission of the persistent source can be shown to be compact
on VLBI angular scales (milliarcseconds), interstellar scintillation
requires structure which is compact on far smaller scales
(microarcseconds; \citealt{mgbh13}). Our ATCA, e-MERLIN and VLBA
observations cannot directly probe the required angular scales.
Hence, the presence of a compact persistent radio source such as an
AGN in WISE\,J071634.59$-$190039.2 allows the scenario proposed by
\citet{wb16} and \citet{aj16}; that the fading radio source coincident
in position with WISE\,J071634.59$-$190039.2 and coincident in time
with FRB\,150418 could be due to interstellar scintillation.

The best probe of the scintillation interpretation will come from
extensive radio photometry measurements, as the signature of
scintillation is well known and can be tested against the data
\citep{aj16}. A careful analysis of all available flux density
measurements of WISE\,J071634.59$-$190039.2 from the ATCA, JVLA,
e-MERLIN, VLBA, EVN, and other facilities should reveal whether the
variability properties of the compact radio source are consistent with
intrinsic AGN variability or scintillation. 

\section*{Acknowledgments}
We thank both e-MERLIN and NRAO for providing observations under
Director's Discretionary Time, and Sarah Burke Spolaor for helping
with the VLBA observations. This paper makes use of software developed
for the Large Synoptic Survey Telescope. We thank the LSST Project for
making their code available as free software at
\url{http://dm.lsstcorp.org}. The National Radio Astronomy Observatory
is a facility of the National Science Foundation operated under
cooperative agreement by Associated Universities, Inc. The Australian
Telescope Compact Array is part of the Australia Telescope National
Facility which is funded by the Commonwealth of Australia for
operation as a National Facility managed by CSIRO.  This work was
based in part on data collected at Subaru Telescope, which is operated
by the National Astronomical Observatory of Japan. CGB acknowledges
support from the European Research Council under the European Union's
Seventh Framework Programme (FP/2007-2013) / ERC Grant Agreement
nr. 337062 (DRAGNET; PI Hessels). SJT is a Western Australian
Premier's Research Fellow. TT was supported by JSPS KAKENHI Grant
Numbers 15K05018 and 40197778. NT is supported by the Toyota
Foundation (D11-R-0830). EK and SB acknowledge the support of the
Australian Research Council Centre of Excellence for All-sky
Astrophysics (CAASTRO), through project number CE110001020. This
research has made use of NASA's Astrophysics Data System.

\bibliographystyle{mnras}

\bsp
\label{lastpage}

\end{document}